\documentclass[fleqn,usenatbib]{mnras}

\usepackage{newtxtext,newtxmath}

\usepackage[T1]{fontenc}

\DeclareRobustCommand{\VAN}[3]{#2}
\let\VANthebibliography\thebibliography
\def\thebibliography{\DeclareRobustCommand{\VAN}[3]{##3}\VANthebibliography}

\usepackage{graphicx}	

\newcommand{\bs}{\boldsymbol}

\title[Metallicity dependence of line-driven winds]{Radiation hydrodynamics simulations of line-driven AGN disc winds: metallicity dependence and black hole growth}

\author[M. Nomura, K. Omukai and K. Ohsuga]{
Mariko Nomura,$^{1,2}$\thanks{E-mail: m-nomura@kure-nct.ac.jp}
Kazuyuki Omukai,$^{2}$
and Ken Ohsuga$^{3}$
\\
$^{1}$Faculty of Natural Sciences, National Institute of Technology (KOSEN), Kure College, 2-2-11 Agaminami, Kure, Hiroshima 737-8506, Japan\\
$^{2}$Astronomical Institute, Graduate School of Science, Tohoku University, 6-3 Aoba, Aramaki, Aoba-ku, Sendai, Miyagi, 980-8578, Japan\\
$^{3}$Center for Computational Sciences, University of Tsukuba, Ten-nodai, 1-1-1 Tsukuba, Ibaraki 305-8577, Japan
}

\date{Accepted 27 July 2021. Received YYY; in original form ZZZ}

\pubyear{2021}

\begin{document}
\label{firstpage}
\pagerange{\pageref{firstpage}--\pageref{lastpage}}
\maketitle

\begin{abstract}
Growth of the black holes (BHs) from the seeds to supermassive BHs (SMBHs, $\sim\!10^9\,M_\odot$) is not understood, but the mass accretion must have played an important role.
We performed two-dimensional radiation hydrodynamics simulations of line-driven disc winds considering the metallicity dependence in a wide range of the BH mass, and investigated the reduction of the mass accretion rate due to the wind mass loss. 
Our results show that denser and faster disc winds 
appear at higher metallicities and larger BH masses.
The accretion rate is suppressed to $\sim\! 0.4$--$0.6$ times the mass supply rate to the disc for the BH mass of $M_{\rm BH}\gtrsim 10^5\,M_{\odot}$ in high-metallicity environments of $Z\gtrsim Z_\odot$, while the wind mass loss is negligible when the metallicity is sub-solar ($\sim 0.1Z_\odot$).
By developing a semi-analytical model,
we found that the metallicity dependence of the line force and the BH mass dependence of the surface area of the wind launch region are the cause of the metallicity dependence ($\propto\! Z^{2/3}$) and BH mass dependencies ($\propto\! M_{\rm BH}^{4/3}$ for $M_{\rm BH}\leq 10^6\,M_\odot$ and $\propto\! M_{\rm BH}$ for $M_{\rm BH}\geq 10^6\,M_\odot$) of the mass-loss rate. 
Our model suggests that the growth of BHs by the gas accretion effectively slows down in the regime $\gtrsim 10^{5}M_{\sun}$ in metal-enriched environments $\gtrsim Z_\odot$. 
This means that the line-driven disc winds may have an impact on late evolution of SMBHs.
\end{abstract}




\section{Introduction}
\label{sec:intro}
Almost all large galaxies harbor supermassive black holes (SMBHs) in their centres.
Recent observations have detected even dozens of SMBHs with mass $\sim\!10^9\,M_\odot$ at redshift $z\gtrsim 6$, or  $\lesssim\!1\,{\rm Gyr}$ after the big bang. 
This early emergence puts strong constraints on their formation scenario (e.g., \citealt{2010AJ....139..906W, 2011Natur.474..616M, 2018Natur.553..473B, 2018ApJS..237....5M, 2020ApJ...897L..14Y}; see also \citealt{2017PASA...34...22G} for review).
To reach such high masses within the short available time, growth from heavy seed BHs of $\sim\!10^3$--$10^5\,M_\odot$ are theoretically preferred  \citep[for review]{2010A&ARv..18..279V, 2013ASSL..396..293H, 2020ARA&A..58...27I}.
In currently favored scenarios, the seeds are supposed to grow via rapid gas accretion close to the Eddington rate. 
However, whether such a high accretion rate is maintained during most of the growth time is still uncertain. 

In active galactic nuclei (AGNs), outflows powered by the accretion flows are thought to be ubiquitous and likely affect growth of the central SMBHs. 
This is supported by recent observations of ultrafast outflows (UFOs) exhibiting huge mass-loss rate and mechanical power.
The UFOs are identified via blueshifted absorption lines of highly ionized iron 
(\ion{Fe}{xxv} and/or \ion{Fe}{xxvi}) found in the X-ray band \citep[e.g.,][]{2002ApJ...579..169C, 2003MNRAS.345..705P}. 
These features are detected in $\sim\!40$ per cent of Seyfert galaxies, indicating that the outflows are likely common in AGNs \citep{2010A&A...521A..57T, 2011ApJ...742...44T, 2012MNRAS.422L...1T, 2013MNRAS.430...60G, 2015MNRAS.451.4169G}.
The typical velocity of the UFOs is $\sim\!0.1$--$0.3c$, where $c$ is the speed of light,
giving an estimate for the mass-loss rate of $\sim\!0.01$--$1\,M_\odot\,{\rm yr^{-1}}$ and the kinetic power of $\sim$(0.1--10 per cent)$L_{\rm Edd}$, where $L_{\rm Edd}$ is the Eddington luminosity \citep{2012MNRAS.422L...1T, 2015MNRAS.451.4169G}.
This kinetic power is large enough to exert the feedback to the host galaxy
\citep{2005Natur.433..604D, 2010MNRAS.401....7H, 2010ApJ...722..642O}, possibly playing a role in the co-evolution of SMBHs and galaxies inferred from the tight correlation between SMBH mass and central velocity dispersion, i.e., the so-called $M-\sigma$ relation
\citep[e.g.,][]{1998AJ....115.2285M, 2000ApJ...539L...9F, 2000ApJ...539L..13G, 2002ApJ...574..740T}.
In addition, the large mass loss might suppress the mass accretion onto the BH directly.

Supposing that the outflow velocity is roughly of the order of the escape velocity from its launching point, 
the velocity of $\sim\!0.1$--$0.3c$ means that the UFOs should be launched from the accretion discs near the SMBHs.
There are a few potential launching mechanisms of such high-velocity disc winds: 
(1) radiation pressure via the electron scattering 
by radiation from super-Eddington accretion flow \citep[e.g.,][]{2009PASJ...61L...7O, 2011ApJ...736....2O},
(2) radiation force due to absorption of ultraviolet (UV) radiation through line transitions of metals \citep[so-called line force, e.g.,][]{2000ApJ...543..686P, 2004ApJ...616..688P, 2016PASJ...68...16N, 2017MNRAS.465.2873N}, 
and (3) magnetic force \citep[e.g.,][]{1982MNRAS.199..883B, 1994ApJ...434..446K, 2015ApJ...805...17F}.
The disc wind accelerated by the line force (line-driven wind) is a promising model.
The line force is 10--1000 times larger than 
that due to Thomson scattering 
when the metal is in a low-ionization state
\citep{1990ApJ...365..321S}, leading to a high velocity disc wind. The observed high-ionization state of the UFOs is apparently inconsistent with the line-driven wind model, but if the wind material becomes highly ionized after it has reached terminal velocity in the low-ionization region near the disc surface, this model can successfully reproduce the absorption features of UFOs \citep{2015MNRAS.446..663H, 2020MNRAS.tmp.3323M}.
The other two models can also explain the high velocity of the outflowing material.  
However, most UFOs are observed in sub-Eddington AGNs, 
in which the electron-scattering force is not powerful enough.
In addition, the magnetic-driven wind model requires an extra mechanism to adjust the ionization state of the wind matter to explain the observed absorption lines. 

Radiation hydrodynamics simulations of the line-driven winds have been developed by 
\citet{2000ApJ...543..686P} and \citet{2004ApJ...616..688P}. 
They clearly showed that the funnel-shaped disc winds are accelerated by the line force 
for the typical parameters of bright AGNs, 
$M_{\rm BH}=10^8\,M_\odot$ and 
$L/L_{\rm Edd}=0.5$, where
$M_{\rm BH}$ and $L$ are the BH mass and the luminosity.
In addition, \cite{2016PASJ...68...16N} and \cite{2017MNRAS.465.2873N} performed radiation hydrodynamics simulations in a wide parameter range ($M_{\rm BH}=10^6$--$10^9\,M_\odot$ and $L/L_{\rm Edd}=0.1$--0.7 ) and found that the line-driven winds well explain the observational features of the UFOs such as the outflow velocity, ionization parameter, column density, 
and mass-loss rate.

However, the simulations so far are not sufficient for studying the SMBH growth because they have assumed a constant mass accretion rate with disc radius. 
\citet[][hereafter N20]{2020MNRAS.494.3616N} developed a new method in which the mass accretion rate is reduced 
at the inner part of the disc in response to the wind mass loss. Those calculations show that the mass accretion rate onto the BH is suppressed to $\sim\! 50$ per cent of the mass supply rate onto the disc for near-Eddington AGNs, with $M_{\rm BH}=10^8\,M_\odot$ and $\dot m_{\rm sup}\sim 0.5$--0.9,
where $\dot m_{\rm sup}=\dot M_{\rm sup}/\dot M_{\rm Edd}$ 
is the mass supply rate $\dot M_{\rm sup}$ 
normalized by the Eddingron rate $\dot M_{\rm Edd}=L_{\rm Edd}/\eta c^2$ with energy conversion rate $\eta=0.06$.
This shows that the mass loss via the line-driven winds significantly affects the mass accretion onto the SMBHs.

The question then arises as to whether the line-driven winds affect the BH growth from seeds to SMBHs. 
Previous simulations of line-driven winds have focused on SMBHs and assumed the solar metallicity. 
However, in order to reveal the role of the line-driven winds in the evolution and the mass accretion processes onto growing seed BHs, the calculations for 
intermediate mass BHs (IMBHs) and investigation of the effect of the metallicity are important.
The metallicity is expected to be sub-solar in the early phase of the BH evolution, and some AGNs are observed to have super-solar metallicites \citep[e.g.,][]{2003ApJ...583..649B, 2006A&A...447..157N}.
The calculations of winds from O stars have shown that the line-driven acceleration is sensitive to the value of the metallicity, because the line force is the radiation force due to the line transitions of metals \citep{2002ApJ...577..389K}. 
The line-driven disc wind around a BH could be largely affected by the expected metal enrichment in the host galaxy. 

In this paper, by improving the hydrodynamics simulations of N20 so as to include the metallicity dependence of the line force, we investigate the effects of the line-driven disc wind on the mass accretion rate in a wide range of the BH mass including IMBHs.

This paper is organized as follows.
In Section \ref{sec:method}, we summarize the improved points of our calculation method. 
Our results and a semi-analytical model explaining the results are presented in Section \ref{sec:result} and \ref{sec:reason}.
Section \ref{sec:summary} is devoted to summary and discussions.

\section{Method}
\label{sec:method}
Our method is almost the same as that of N20,
which basically follows the calculations of \citet{2000ApJ...543..686P} and \citet{2004ApJ...616..688P} 
but is modified to satisfy the conservation of the total mass of the accretion disc and winds. 
See N20 for details of the difference between our method and that of the previous works.
In this paper, we modify the computational domain and the functional form of the force multiplier
so as to include the metallicity dependence of the line force and apply our calculation method to a wider range of the BH mass.
Here, we briefly explain the outline of our method and describe the difference from N20.

We employ the spherical polar coordinate $(r,\theta,\varphi)$
, where $r$ is the distance from the origin of the coordinate, $\theta$ is the polar angle, and $\varphi$ is the azimuthal angle. 
The simulations are performed in two-dimensions assuming the axial symmetry with respect to the rotation axis of the disc. 
We calculate the following basic equations of the hydrodynamics, i.e.,  
the equation of continuity,
\begin{equation}
\frac{\partial \rho}{\partial t}+\nabla\cdot(\rho \mbox{\boldmath $v$})
=0,
\label{eoc}
\end{equation}
the $r$, $\theta$, and $\varphi$ components of the equations of motion,
\begin{equation}
\frac{\partial (\rho v_r)}{\partial t}+\nabla\cdot(\rho v_r \mbox{\boldmath $v$})
=-\frac{\partial p}{\partial r}+\rho\Bigg[\frac{v_\theta^2}{r}+\frac{v_\varphi^2}{r}+g_r+f_{{\rm rad},\,r}\Bigg],
\label{eom1}
\end{equation}
\begin{equation}
\begin{split}
\frac{\partial (\rho v_\theta)}{\partial t}+&\nabla\cdot(\rho v_\theta \mbox{\boldmath $v$})\\
&=-\frac{1}{r}\frac{\partial p}{\partial \theta}+\rho\Bigg[-\frac{v_r v_\theta}{r}+\frac{v_\varphi^2}{r}\cot \theta+g_\theta+f_{{\rm rad},\,\theta}\Bigg],
\label{eom2}
\end{split}
\end{equation}
\begin{equation}
\frac{\partial (\rho v_\varphi)}{\partial t}+\nabla\cdot(\rho v_\varphi \mbox{\boldmath $v$})
=-\rho\Bigg[\frac{v_\varphi v_r}{r}+\frac{v_\varphi v_\theta}{r}\cot \theta\Bigg],
\label{eom3}
\end{equation}
and the energy equation,
\begin{equation}
\frac{\partial}{\partial t}\Bigg[  \rho \Bigg(\frac{1}{2}v^2+e\Bigg) \Bigg]
+\nabla \cdot \Bigg[  \rho\mbox{\boldmath $v$} \Bigg(\frac{1}{2}v^2+e+\frac{p}{\rho}\Bigg) \Bigg]
=\rho\mbox{\boldmath $v$}\cdot\mbox{\boldmath $g$}+\rho\mathcal L,
\label{eoe}
\end{equation} 
where $\rho$ is the mass density,
\mbox{\boldmath $v$}$=(v_r,\,v_\theta,\,v_\varphi)$
is the velocity,
$p$ is the gas pressure, $e$ is the internal energy per unit mass and
\mbox{\boldmath $g$}$=(g_r,\,g_\theta)$ 
is the gravitational acceleration of the BH.
The equation of state $p/\rho=(\gamma -1)e$ with the adiabatic component $\gamma=5/3$ 
is employed.
In the last term of Eq.\ref{eoe}, 
$\mathcal L$ is the net cooling rate including Compton heating/cooling, 
X-ray photoionization heating, recombination cooling, 
bremsstrahlung cooling, and line cooling (see also N20).

In Eqs.\ref{eom1} and \ref{eom2}, 
$\bs f_{\rm rad}=(f_{{\rm rad},\,r},\,f_{{\rm rad},\theta} )$ is the radiation force described as 
\begin{equation}
\bs f_{\rm rad}=\frac{\sigma_{\rm e} \bs F_{\rm D}}{c}+\frac{\sigma_{\rm e} \bs F_{\rm line}}{c}M,
 \label{radforce}
\end{equation}
where $\sigma_{\rm e}$ is the mass-scattering coefficient for free electrons,
$\bs F_{\rm D}$ is the radiation flux emitted from the accretion disc
integrated by the wavelength throughout the entire range, 
and $\bs F_{\rm line}$ is the line-driving flux, 
which is the same as $\bs F_{\rm D}$ but integrated across the UV band of $200$--$3200$\,\AA.
The second term of Eq. \ref{radforce} is the line force.
As mentioned in Section \ref{sec:intro}, 
the line transitions depend on the wavelength of the radiation.
The line force is exerted mainly by the radiation flux in the UV band (200--3200\,\AA),
because the metal lines are densely distributed \citep[e.g.,][]{1975ApJ...195..157C, 1990ApJ...365..321S}.
Thus, in this paper, we evaluate the line force using the UV (200--3200\,\AA) radiation flux
and the corresponding force multiplier $M$ same as \citet{2004ApJ...616..688P}. 
Here $M$ is the numerical factor indicating how much the spectral lines enhance the radiation force compared to the Thomson scattering.
The radial components of the fluxes are estimated as $ F^r_{\rm D}= F^r_{\rm D,\,thin}e^{-\tau_{\rm e}}$ and $ F^r_{\rm line}= F^r_{\rm line,\,thin}e^{-\tau_{\rm e}}$ respectively,
where $\tau_{\rm e}$ is the electron-scattering optical depth estimated as 
$\tau_{\rm e} =\int^r_{r_{\rm in}} \sigma_{\rm e} \rho(r',\,\theta) dr'$,
where $r_{\rm in}$ is the inner boundary of the computational box.
We ignore the attenuation in the $\theta$-direction as 
$ F^\theta_{\rm D}= F^\theta_{\rm D,\,thin}$ and $ F^\theta_{\rm line}= F^\theta_{\rm line,\,thin}$.
We calculate $\bs F_{\rm D,\,thin}$ and $\bs F_{\rm line,\,thin}$ by integrating intensity transferred in the optically-thin media from the grids on the disc to the point of interest.
Here we employ the standard disc model \citep{1973A&A....24..337S}.
We divide the hot region of the disc where the effective temperature is larger than $3\times 10^3\,{\rm K}$ into the grids.
In contrast to the previous method where we prepared 4096 uniform grids both in the $r$- and $\varphi$-directions, we here prepare 12800 grids whose sizes are determined by $\Delta r_i /\Delta r_{i-1} =1.0005$ in the $r$-direction. In the $\varphi$-direction, we set 1600 uniform grids in the range of $0\leq \varphi <2\pi$.
In order to resolve the hot region of the disc ($T_{\rm eff}>3\times 10^3 \,{\rm K}$) 
for the IMBHs
($M_{\rm BH}=10^3$--$10^6 \,M_\odot$), 
a large number of grids is required in the  $r$-direction.
This is because the size of the hot region normalized by Schwarzschild radius $R_{\rm S}$ 
increases with the decrease of the BH mass.  
For example, the outer radius of the hot region is $\sim\! 1000R_{\rm S}$ for $M_{\rm BH}=10^8\,M_\odot$ while it reaches as far as $3\times 10^4 R_{\rm S}$ for $M_{\rm BH}=10^3\,M_\odot$. 

In the second term of Eq.\ref{radforce}, $M$ is the force multiplier,
which is a function of the local optical depth parameter,
\begin{equation}
  t=\sigma_{\rm e}\rho v_{\rm th}\left| \frac{dv}{ds} \right|^{-1},
  \label{eq-t}
\end{equation}
the ionization parameter,
\begin{equation}
\xi=\frac{4\pi F_{\rm X}}{n},
\end{equation}
and the metallicity, $Z$,
where $v_{\rm{th}}$ is the thermal speed of hydrogen gas
whose temperature is $25,000\,{\rm K}$ ($v_{\rm th}=20\,{\rm km\,s^{-1}}$),
$dv/ds$ is the velocity gradient along the light-ray,
$n$ is the number density, 
and $F_{\rm X}$ is the X-ray flux
from the central source.  
We set the X-ray irradiating source as a point source located at the centre of the coordinate with the luminosity $L_{\rm X}=0.1\eta \dot{M}_{\rm sup}c^2$ in the same manner as N20.
The X-ray flux is estimated as 
$F_{\rm X} = L_{\rm X}e^{-\tau_{\rm X}}/4\pi r^2$.
Here, using the mass extinction coefficient with a simple assumption that 
$\sigma_{\rm X}=\sigma_{\rm e}$ for $\xi \geq 10^5$ or $\sigma_{\rm X}=100\sigma_{\rm e}$ for $\xi < 10^5$,
we estimate the optical depth for the X-ray as 
$\tau_{\rm X} =\int^r_{r_{\rm in}} \sigma_{\rm X}(\xi) \rho(r',\,\theta) dr'$.
In this paper, we do not consider the X-rays from the accretion disc surface 
so as to understand $M_{\rm BH}$-dependence of the line-driving mechanism with a simplified model, but  
in Section \ref{sec:summary}, we discuss effects of the X-ray emitted from the disc on the results.

\begin{figure}
 \begin{center}
  \includegraphics[width=\columnwidth]{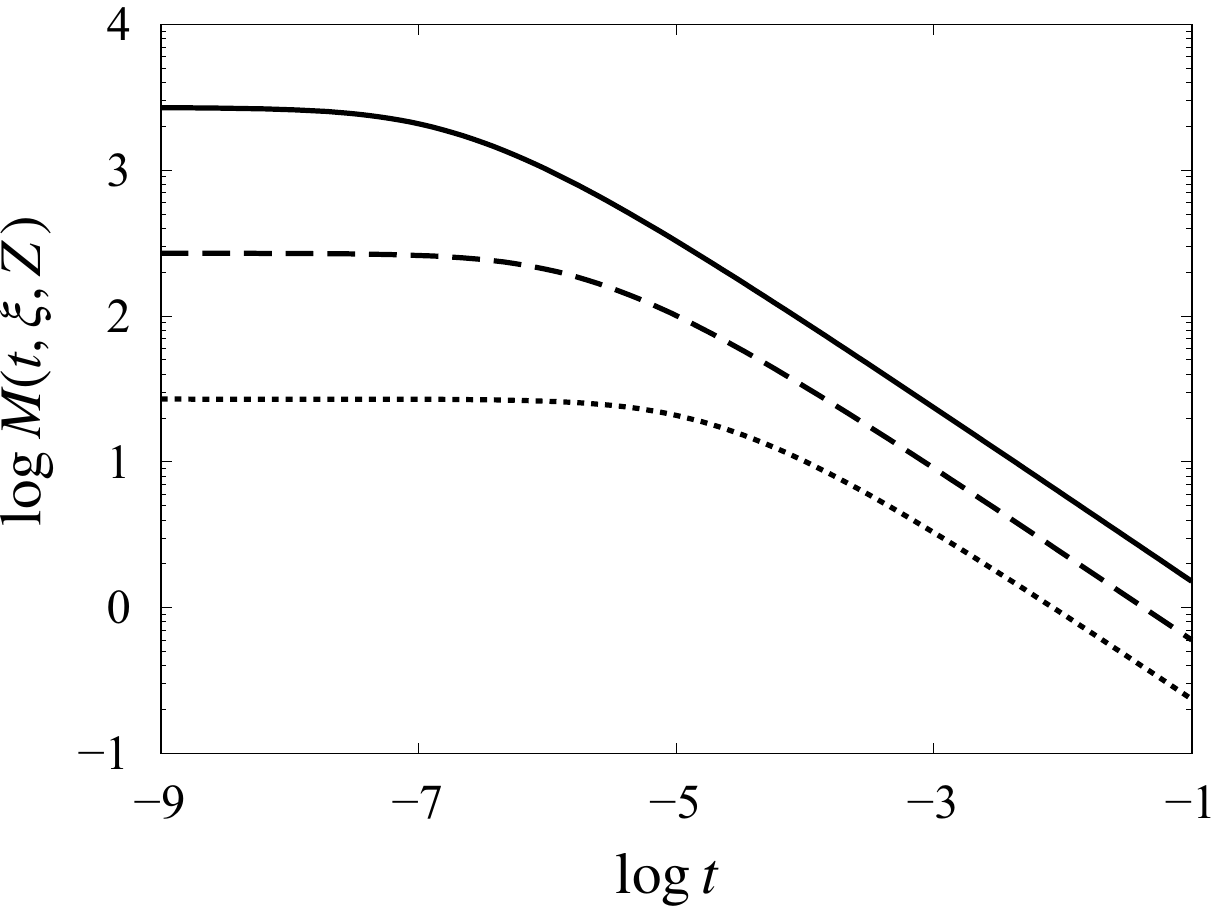}
 \end{center}
 \caption{
Force multiplier as a function of the local optical depth parameter. The ionization parameter is fixed at $\log \xi =-2$. 
Solid, dashed, and dotted lines show the force multipliers for $Z/Z_\odot=1$, 0.1, and 0.01, respectively.} 
 \label{fig1}
\end{figure}

The metallicity dependence of the force multiplier 
has been discussed by \citet{2002ApJ...577..389K} in the context of winds from O stars.
According to this work, by assuming that the line strength is proportional to the metallicity, we introduce the force multiplier described as
\begin{equation}
 M(t,\xi,Z) =
  \left( \frac{Z}{Z_\odot}\right) ^{1-\alpha} Kt^{-\alpha}\left[ \frac{\{1+(Z/Z_\odot)
   \eta_{\rm max}t\}^{1-\alpha}-1}{\{(Z/Z_\odot)
   \eta_{\rm max}t\}^{1-\alpha}}\right ],
   \label{forceM}
\end{equation}
where $\alpha$ is set to 0.6, 
$Z_\odot$ is the solar metallicity, 
$K$ and $\eta_{\rm max}$ are functions of the ionization parameter written as 
\begin{equation}
 K=0.03+0.385\exp (-1.4\xi^{0.6}),
\end{equation}
and 
\begin{equation}
\log \eta_{\rm max}=
\left\{
    \begin{array}{l}
     6.9\exp(0.16\xi^{0.4}) \qquad \log\xi \leq 0.5 \\
     9.1\exp(-7.96\times 10^{-3}\xi) \qquad \log\xi > 0.5
    \end{array} 
\right.
 .
\end{equation}
When we assume $Z=Z_\odot$, Eq.\ref{forceM} agrees with the force multiplier presented by \citet{1990ApJ...365..321S}, which was used in N20.
Fig. \ref{fig1} 
demonstrates the force multiplier defined by 
Eq.\ref{forceM} as a function of $t$ for fixed ionization parameter $\log \xi =-2$ and three different metallicities $Z/Z_\odot=1$ (solid line), $0.1$ (dashed line), and $0.01$ (dotted line). 
For small $t$ ($\log t \lesssim -6$), where the lines are optically thin,
the force multiplier is proportial to the metallicity as 
$\left(1-\alpha\right)\left(Z/Z_\odot \right)K\eta_{\rm max}^{\alpha}$. This is because almost all lines contribute to the line acceleration. 
For large $t$ ($\log t \gtrsim -6$), where some lines become optically thick, the force multiplier is reduced to $\left( Z/Z_\odot \right)^{1-\alpha}Kt^{-\alpha}$ because optically thick lines are less efficient in accelerating the wind.
This metallicity dependence agrees with the results of \citet{1982ApJ...259..282A} and \citet{2002ApJ...577..389K}.

For compulational reasons, we divide the mass range of the central BHs in the intermediate ($\leq 10^{6} M_{\odot}$) and supermassive ($\geq 10^{6} M_{\odot}$) ranges and use different sizes of the computational box.
For the IMBHs ($M_{\rm BH} \leq 10^6 \,M_\odot$),
the computational domain is set to $r_{\rm in}\leq r\leq r_{\rm out}$ 
and $0\leq \theta \leq 90^\circ$.
We set $r_{\rm in}$ and $r_{\rm out}$ so that the computational domain includes the UV-bright region of the disc. Based on the standard disc model \citep{1973A&A....24..337S}, the disc radius is a
function of $M_{\rm BH}$, $\dot{m}_{\rm sup}$, 
and the effective temperature, $T_{\rm eff}$, as
\begin{equation}	
		\label{radii}
		\frac{r}{R_{\rm S}}
		=\left(\frac{3c^5}{16\sigma_{\rm e}\sigma G}\right)^{1/3}
  T_{\rm eff}^{-4/3}\dot m_{\rm sup}^{1/3}M_{\rm BH}^{-1/3},
	\end{equation}
where $\sigma$ and $G$ are Stefan-Boltzmann constant and the gravitational constant.
At the inner and outer radii, we set the effective temperatures to $2.5\times 10^5\,{\rm K}$ and $1.6\times 10^4\,{\rm K}$, respectively. In this temperature range, the radiation has a peak of the blackbody spectra within the UV band of $200$--$3200$\,\AA. Eq.\ref{radii} indicates that the location of the UV-bright region is shifted outward as the BH mass decreases and the normalised mass supply rate increases.
For larger BH masses ($M_{\rm BH}\geq 10^6 \,M_\odot$), we employ the same size of computational domain as in N20, $r_{\rm in}=30R_{\rm S}$ and $ r_{\rm out}=1500 R_{\rm S}$. 
We assume that the matter located within $<30 R_{\rm S}$ is highly ionized by hard X-ray coronal radiation from the central region in the case of AGNs.

We employ the mass-conserving iterative method (see Section 2.3 in N20 for details), in which we take into account the reduction in mass accretion rate through the disc caused by the launching of the wind.
The boundary conditions and initial conditions are the same as in N20.

\section{Results}
\label{sec:result}
\subsection{Overview of metallicity and BH mass dependence}

\begin{figure*}
 \begin{center}
  \includegraphics[width=\textwidth]{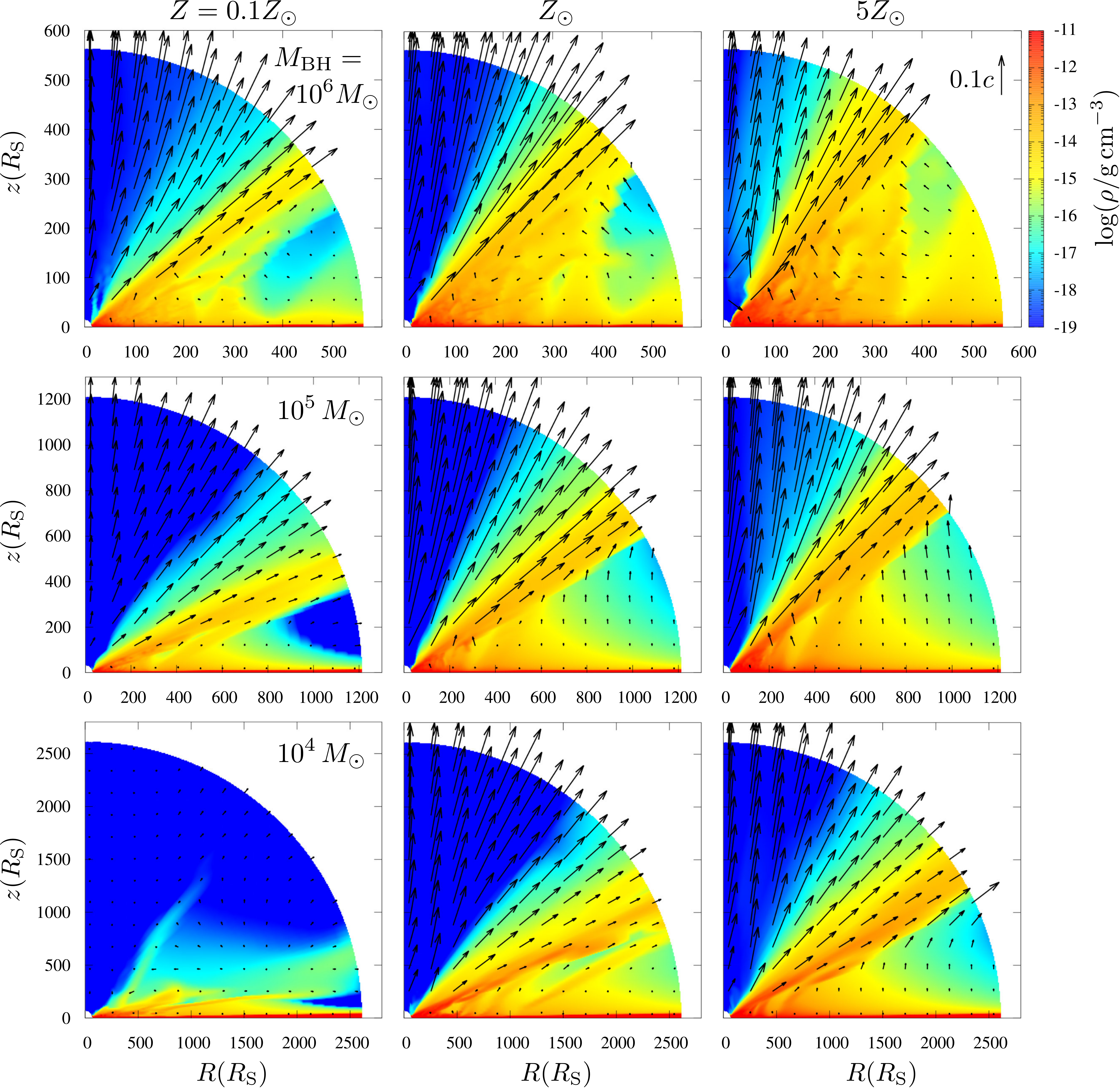}
 \end{center}
 \caption{Time averaged density map and velocity structure in the $R$-$z$ plane for $\dot m_{\rm sup}=0.5$.
   Left, middle, and right columns show the resuls for $Z=0.1Z_\odot$, $Z_\odot$, and $5Z_\odot$, respectively.
   Top, middle, and bottom rows correspond to $M_{\rm BH}=10^6\,M_\odot$, $10^5\,M_\odot$, and $10^4\,M_\odot$. 
   In each panel, the $z$-axis is the rotational axis of the disc and $R$ is the distance from the $z$-axis. 
   Radial scales are 15--580$R_{\rm S}$, 32--1240$R_{\rm S}$, and 68--2680$R_{\rm S}$ for $M_{\rm BH}=10^6\,M_\odot$, $10^5\,M_\odot$, and $10^4\,M_\odot$.
   Note the different spatial scales among the models.
   }
 \label{fig2}
\end{figure*}

First we overview the metallicity and BH mass dependence of the line-driven winds
with $\dot m_{\rm sup}=0.5$.
Fig. \ref{fig2} shows the time averaged density map and velocity structure in the $R$-$z$ plane across the parameter space of
$0.1Z_\odot \leq Z\leq 5Z_\odot$ and $10^4\, M_\odot \leq M_{\rm BH}\leq 10^6\, M_\odot$.
The left, middle, and right columns correspond to $Z=0.1Z_\odot$, $Z_\odot$ and $5Z_\odot$.
The top, middle, and bottom rows show the results for $M_{\rm BH}=10^6\,M_\odot$, $10^5\,M_\odot$, and $10^4\,M_\odot$, respectively.
In each panel, the $z$-axis is the rotational axis of the disc and $R$ is the distance from the $z$-axis. 
The radial ranges of the computational domains are 15--580$R_{\rm S}$, 32--1240$R_{\rm S}$, and 68--2680$R_{\rm S}$ for $M_{\rm BH}=10^6\,M_\odot$, $10^5\,M_\odot$, and $10^4\,M_\odot$, respectively.

We can find denser and faster winds in the upper right panels with higher metallicity and BH mass.
In each row, we can clearly see that denser and faster winds are launched from the disc
toward (vertically) upper directions in higher metellicity cases.
This is caused by the larger value of the force multiplier at higher metallicities as shown in Fig. \ref{fig1}
(see Section \ref{sec:reason} for details).
Each column shows that the outflow becomes denser and faster with increasing BH mass.
This tendency can be explained by the $M_{\rm BH}$-dependence of
the surface area of the wind launching region:
more massive BHs have larger UV-bright regions (see Section \ref{sec:reason} for details).

\subsection{Metallicity and BH mass dependence of mass accretion rate}

\begin{figure}
 \begin{center}
  \includegraphics[width=\columnwidth]{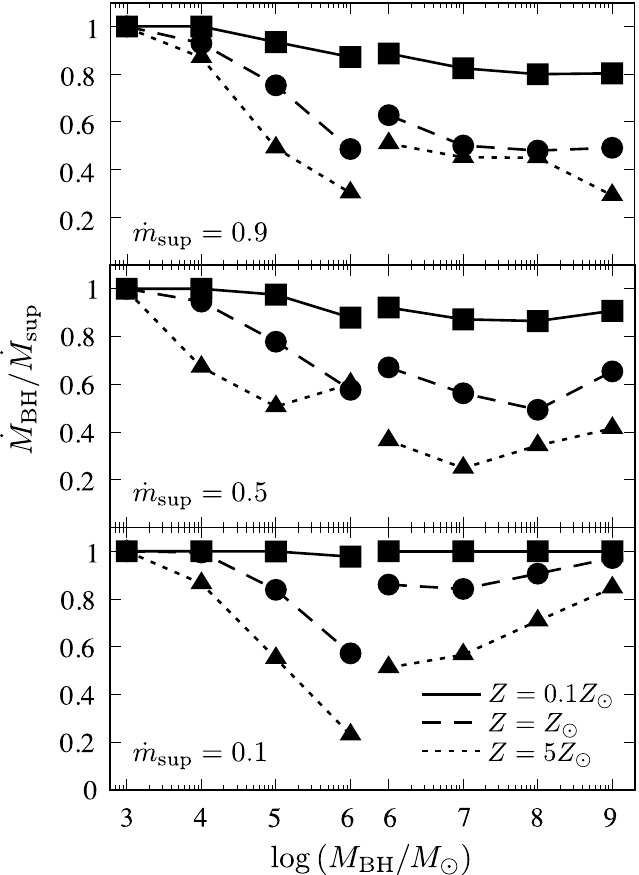}
 \end{center}
 \caption{Mass accretion rate normalized by the mass supply rate as a function of the BH mass for
$\dot m_{\rm sup}=0.9$ (top panel), 0.5 (middle panel), and 0.1 (bottom panel). The solid, dashed, and dotted lines in each panel show the results for $Z=0.1Z_\odot$, $Z_\odot$, and $5Z_\odot$, respectively.} 
 \label{fig3}
\end{figure}

Here we focus on how the mass accretion rate $\dot M_{\rm BH}$ depends on the metallicity
and the BH mass.
Fig. \ref{fig3} shows the mass accretion rate normalized
by the mass supply rate as a function of the BH mass when we employ
$\dot m_{\rm sup}=0.9$ (top panel), 0.5 (middle panel), and 0.1 (bottom panel).
For the IMBHs ($M_{\rm BH}\leq 10^6\,M_\odot$,
left hand side plots in Fig. \ref{fig3}),
we employ the radial computational domain covering the UV-bright region,
while we use the fixed radial computational domain, $30R_{\rm S}\leq r\leq 1500R_{\rm S}$, for larger BH masses
($M_{\rm BH}\geq 10^6\,M_\odot$, right hand side plots in Fig. \ref{fig3}). 

In all three panels,
the normalized mass accretion rate decreases in the order of
$Z/Z_\odot=0.1$ (solid line), 1 (dashed line), and 5 (dotted line) regardless of the BH mass.
In the IMBH mass range
($M_{\rm BH}\leq 10^6\,M_\odot$), each function shows that
the normalized mass accretion rate is reduced
by the increase of the BH mass.
These plots are consistent with the result
that the density and velocity increase
with the growth of the metallicity and the BH mass (Fig. \ref{fig2}).
In the SMBH range ($M_{\rm BH}\geq 10^6\,M_\odot$),
the normalized mass accretion rate is almost constant
with the changes of the BH mass except for the case of $\dot m_{\rm sup}=0.1$.

For $M_{\rm BH}\leq 10^6\,M_\odot$,
we find that, at given BH mass and metallicity, the normalized mass accreton rate  is almost the same regardless of $\dot{m}_{\rm sup}$.
In each panel, the normalized mass accretion rate is $\sim\! 1$ for $M_{\rm BH}=10^3\,M_\odot$ even for super-solar metallicity, because the mass-loss rate of the winds $\dot M_{\rm out}$, if any, is quite small $\dot M_{\rm out}/\dot M_{\rm sup}\lesssim 5$ per cent.
When $M_{\rm BH}=10^4\,M_\odot$, the normalized mass accretion rate is close to $1$ for $Z\leq Z_\odot$ and $\sim\! 0.7$--$0.9$ for $Z=5Z_\odot$.
This means that almost all supplied mass accretes onto the central BH,
and the growth of the BH is not suppressed by the line-driven disc winds
for $M_{\rm BH}\lesssim 10^4\,M_\odot$.
For $M_{\rm BH}=10^5\,M_\odot$, the normalized mass accretion rate is still large $\gtrsim 0.8$ for $Z\leq Z_\odot$, while $\dot M_{\rm BH}/\dot M_{\rm sup}\sim 0.5$ for $Z=5Z_\odot$. 
For $M_{\rm BH}=10^6\,M_\odot$ corresponding to an AGN with relatively low BH mass, the normalised mass accretion rate is less than $\sim \! 0.6$ except for $Z=0.1Z_\odot$ ($\dot M_{\rm BH}/\dot M_{\rm sup}\gtrsim 0.9$). To summarize, the line-driven winds may suppress the mass accretion for $M_{\rm BH}\gtrsim 10^5\,M_\odot$ in high metallicity environments.

For the larger BH mass ($M_{\rm BH}\geq 10^6\,M_\odot$),
the normalized mass accretion rate remains almost constant
with respect to the BH mass
in the top and middle panels ($\dot m_{\rm sup}=0.5$ and 0.9), 
with its value
$\sim\! 0.8$--$0.9$, $\sim\! 0.5$--$0.7$, and $\sim\! 0.3$--$0.5$
for $Z=0.1Z_\odot$, $Z_\odot$, and $5Z_\odot$, respectively.
These plots suggest that the line-driven winds suppress to some extent the mass accretion onto the SMBHs in high metallicity and high mass supply rate environments. 
In the bottom panel ($\dot m_{\rm sup}=0.1$),
$\dot M_{\rm BH}/\dot M_{\rm sup}$ increases gradually
with the growth of the BH mass. When $M_{\rm BH}=10^9\,M_\odot$, we find $\dot M_{\rm BH}/\dot M_{\rm sup}\gtrsim 0.8$ even for $Z=5Z_\odot$.
The temperature of the discs is too low to emit strong line-driving UV radiation
for $\dot m_{\rm sup}=0.1$ and $M_{\rm BH}\gtrsim 10^8\,M_\odot$.
This would be the reason for 
the smaller mass-loss late of the winds,
leading the large mass accretion rate.

Here we remark difference in computational domains.
The disc wind is thought to be launched from the 
UV-bright region of the disc.
However, the region very close to the BH is filled with 
ionizing X-ray radiation, 
which may prevent the generation of the wind.
Therefore, in this study, 
we set the computational domain 
to cover the UV-bright region with $r\geq30R_{\rm S}$.
In the smaller $M_{\rm BH}$ regime, 
we adopt the computational domain that includes the whole UV-bright region with $T_{\rm eff}=1.6\times 10^4$--$2.5\times 10^5\,{\rm K}$. This is because the UV-bright region is relatively located outside since the disc temperature tends to be high. 
For the larger $M_{\rm BH}$ regime, the computational domain is set to be $30-1500R_{\rm S}$, although the UV-bright region extends into the region of $r<30R_{\rm S}$ due to the relatively low disc temperature.

In this paper, $r_{\rm in}$ is set to $30R_{\rm S}$
so that $M_{\rm BH}$ of about $10^6\,M_\odot$ seems reasonable to switch between the above two computational domains.
In Fig. \ref{fig3}, we compare the mass accretion rate 
calculated with two different methods for $M_{\rm BH}=10^6\,M_\odot$, and confirm that the difference between the results was less than or equal to a factor of two.
The reason why the discrepancy is relatively large for $\dot m_{\rm sup}=0.1$ is 
that the effective temperature of $\dot m_{\rm sup}=0.1$ is lower than that of $\dot m_{\rm sup}\geq 0.5$.
Since most of the UV emission region is inside $r_{\rm in}=30R_{\rm S}$ (i.e., outside the simulation box for the SMBH method), 
launching of the disc wind is suppressed for SMBH method. 
On the other hand, simulations with the computationl domain for IMBH, 
the matter is ejected as the wind from inside 30$R_{\rm S}$.

The $r_{\rm in}$ used in this study would correspond to the size of the X-ray corona in reality. Although the size of the X-ray corona is still quite uncertain, it is expected to be $\sim\! 10-100R_{\rm S}$ \citep[e.g.,][]{2016A&A...594A..71E,2018MNRAS.480.1247K}, which is not so deviated from the setting of the present study ($r_{\rm in} = 30R_{\rm S}$). Even if we change $r_{\rm in}$ slightly to the extent that the observations suggest, our results will not change so much. Thus,
it is plausible that the declining tendency of the mass accretion rate with the BH growth would stop once the BH mass enters the SMBH regime, i.e., $\gtrsim 10^6\,M_\odot$.
We leave the effect of corona geometry to a future study.

Reduction in the mass accretion rate causes inward shift of the UV-bright region in the disc
(see Eq. \ref{radii}),
but we do not take into account this effect when we set the computational box.
The UV-bright region
($1.6\times 10^4\,{\rm K} \leq T_{\rm eff} \leq 2.5\times 10^5\, {\rm K}$)
of the disc with the non-reduced mass accretion rate
$\dot M_{\rm BH}=\dot M_{\rm sup}$
is entirely covered by the computational box in our model.
The line-driving UV luminosity emitted
within the computational box for the discs having $\dot M_{\rm BH}<\dot M_{\rm sup}$
could become smaller than that esimated for $\dot M_{\rm BH}=\dot M_{\rm sup}$.
However, in the parameter space in this study,
the difference is not so large and would not affect the dynamics of the winds.
In the case of $\dot m_{\rm sup}=0.1$, $M_{\rm BH}=10^6\,M_\odot$, and $Z=5Z_\odot$ showing the smallest $\dot M_{\rm BH}/\dot M_{\rm sup}$, the inner boundary of the computational box is $8.6R_{\rm S}$.
The reduced mass accretion rate normalized by the Eddington rate is $\dot m_{\rm BH}=\dot M_{\rm BH}/\dot M_{\rm Edd}=0.023$, and the UV-bright region (the radius where $T_{\rm eff}=2.5\times 10^5\,{\rm K}$) shifts to $5.3R_{\rm S}$.
The line-driving luminosity emitted within the current computational box
$8.6R_{\rm S}\leq r\leq 330R_{\rm S}$ and
that emitted within the shifted UV-bright region $5.3R_{\rm S}\leq r\leq 330R_{\rm S}$ are
$\sim\! 1.6\times 10^{42}\,{\rm erg\,s^{-1}}$ and $\sim\! 2.1\times 10^{42}\,{\rm erg\,s^{-1}}$ respectively.
The current computational box covers $\sim\! 78$ per cent of the shifted UV-bright region and this difference would not largely affect the results.

\section{Semi-analytical modeling for the mass-loss rate}
\label{sec:reason}

In this section, we introduce one-dimensional semi-analytical model of the disc wind, which explains the reason for $M_{\rm BH}$- and $Z$-dependence of the mass-loss rate. In this model, we modify the model of \citet[][hereafter CAK75]{1975ApJ...195..157C} and apply it to the disc wind near the disc surface where the wind is accelerated in the direction of the nearly $z$-axis.

The equations that regulate the steady structure of the disc wind
near the disc surface are the mass conservation,
\begin{equation}
  \label{eq-mc}
  \dot M_{\rm out}=S\rho v_z,
\end{equation}
and the equation of motion,
\begin{equation}
  \label{eq-eom}
v_z \frac{d v_z}{d z}=-\frac{GM_{\rm BH}z}{(R_{\rm l}^2+z^2)^{3/2}}+\frac{\sigma_{\rm e}}{c}\sigma T_{\rm eff}(R_{\rm l})^4(1+M)-\frac{1}{\rho}\frac{dp}{dz},
\end{equation}
where $S$ is the surface area of the launching region of the disc, 
$v_z$ is the $z$-component of the velocity, 
and $T_{\rm eff}(R_{\rm l})$ is the effective temperature of the disc at the launching radius $R=R_{\rm l}$.
We assume the isothermal equation of $p=\rho c_{\rm s}^2$ with the sound speed $c_{\rm s}$.
Here, the force multiplier is 
\begin{equation}
  \label{eq-fm}
M=K\left( Z/Z_\odot \right)^{1-\alpha}t^{-\alpha},
\end{equation}
where $t$ is the local optical depth parameter given by Eq. \ref{eq-t}.
In this semi-analytical model focusing on a weakly ionized launching region,
we ignore the $\xi$-dependence of the force multiplier and $K$ is assumed to be constant, $K=0.4$.

From Eqs. \ref{eq-mc}--\ref{eq-fm}, we can derive the equation for $v_z$,
\begin{equation}
  \begin{split}
  \label{eq-velocity}
  &\left( v_z -\frac{c_{\rm s}^2}{v_z} \right)
  \frac{dv_z}{dz}=-\frac{GM_{\rm BH}z}{(R_{\rm l}^2+z^2)^{3/2}}+\frac{\sigma_{\rm e}}{c}\sigma T_{\rm eff}(R_{\rm l})^4 \\  
  &\ \ \ \ \ \ +\frac{\sigma_{\rm e}}{c}\sigma T_{\rm eff}(R_{\rm l})^4
  K\left( \frac{Z}{Z_\odot} \right)^{1-\alpha} 
  \left( \frac{S}{\sigma_{\rm e} v_{\rm th} \dot M_{\rm out}} \right)^{\alpha} 
  \left( v_z \frac{dv_z}{dz}\right)^\alpha,
  \end{split}
\end{equation}
which corresponds to Eq. 20 in CAK75.
Following CAK75 approach, we introduce new variables, $w$ and $h$, defined by
\begin{equation}
  w=\frac{1}{2}v_z^2,
\end{equation}
and
\begin{equation}
h(z)=-\frac{GM_{\rm BH}z}{(R_{\rm l}^2+z^2)^{3/2}}+\frac{\sigma_{\rm e}}{c}\sigma T_{\rm eff}(R_{\rm l})^4.
\end{equation}
Since the mass-loss rate $\dot M_{\rm out}$ and the effective temperature $T_{\rm eff}(R_{\rm l})$ are constant along the $z$-direction, we define the constant $C$ as
\begin{equation}
  \label{eq-C}
  C=\frac{\sigma_{\rm e}}{c}\sigma T_{\rm eff}(R_{\rm l})^4
  K\left( \frac{Z}{Z_\odot} \right)^{1-\alpha} 
  \left( \frac{S}{\sigma_{\rm e} v_{\rm th} \dot M_{\rm out}}\right)^{\alpha}.
\end{equation}
Using $w$, $h$, and $C$,
we can rewrite Eq. \ref{eq-velocity} as
\begin{equation}
  \label{eq-w'}
  F(z,w,w')=\left( 1-\frac{c_{\rm s}^2}{2w} \right)w'-h(z)-C(w')^\alpha=0,
\end{equation}
where $w'$ stands for $dw/dz$.
Eq. \ref{eq-w'} is the same as Eq. 26 in CAK75, except that we use $z$
instead of $u=1/r$.

For given $z$ and $w$, Eq. \ref{eq-w'} is the equation of $w'$, whose number of solutions depends on the values of $z$, $w$, $h(z)$ and $C$.
According to CAK 75, the cases are:
\renewcommand{\labelenumi}{(\Roman{enumi})}
\begin{enumerate}
\item for $w<c_{\rm s}^2/2$ and $h<0$, there is one solution for $w'$.

\item for $w>c_{\rm s}^2/2$, $h<0$ and \\
  \hfill$1-c_{\rm s}^2/2w<\alpha(1-\alpha)^{(1-\alpha)/\alpha}C^{1/\alpha}(-h)^{-(1-\alpha)/\alpha}$, \\
  \hfill
  there are two solutions for $w'$.

\item for $w>c_{\rm s}^2/2$ and $h>0$, there is one solution for $w'$.

\item for $w>c_{\rm s}^2/2$, $h<0$ and \\
  \hfill$1-c_{\rm s}^2/2w>\alpha(1-\alpha)^{(1-\alpha)/\alpha}C^{1/\alpha}(-h)^{-(1-\alpha)/\alpha}$, \\
  \hfill
  there is no solution.
  
\item for $w<c_{\rm s}^2/2$ and $h>0$, there is no solution.
\end{enumerate}
We find one subsonic solution and one supersonic solution in the cases I and I\hspace{-.1em}I\hspace{-.1em}I, respectively.
In the $z$-$w$ plane, the subsonic branch in the region I is connected to the supersonic branch in the region I\hspace{-.1em}I\hspace{-.1em}I 
at the border between the regions I\hspace{-.1em}I and I\hspace{-.1em}V,
where the local minimum of the function $F(z,w,w')$ is zero.
This point is called the singular point in CAK75 and defined by Eq. \ref{eq-w'} and the condition of  
\begin{equation}
  \label{eq-df}
  \frac{\partial F(z,w,w')}{\partial w'}
  =1-\frac{c_{\rm s}^2}{2w}-\alpha C(w')^{\alpha-1}=0.
\end{equation}

Additionally, if we suppose that $w'$ is smooth throughout the wind branch,
we can define $w^{\prime\!\prime}=dw'/dz$ and find $\partial F/\partial z+w'(\partial F/\partial w)+w^{\prime\!\prime}(\partial F/\partial w')=0$ by differentiating Eq. \ref{eq-w'} with respect to $z$. At the singular point, where $\partial F/\partial w'=0$, we obtain
\begin{equation}
  \label{eq-df2}
  \frac{\partial F}{\partial z}+w'\frac{\partial F}{\partial w}=0.
\end{equation}
The wind property at the singular point is determined by Eqs. \ref{eq-w'}--\ref{eq-df2}. 
If the height of the singular point $z=z_{\rm c}$ is given, $w$, $w'$, and $C$ including $\dot M_{\rm out}$ can be derived from these equations.

From Eqs. \ref{eq-w'}--\ref{eq-df2}, and the deficition of $C$ (Eq. \ref{eq-C}), we have 
the mass-loss rate as 
\begin{equation}
  \begin{split}
  \label{eq-mout}
  \dot M_{\rm out}=
  &\frac{S}{\sigma_{\rm e}v_{\rm th}}
  \alpha (1-\alpha)^{(1-\alpha)/\alpha}K^{1/\alpha}\left( \frac{Z}{Z_\odot}\right)^{(1-\alpha)/\alpha}\\
  &\times\left( g_{\rm c}^z-\frac{\sigma_{\rm e}}{c}\sigma T_{\rm eff}(R_{\rm l})^4 \right)^{-(1-\alpha)/\alpha}\left( \frac{\sigma_{\rm e}}{c}\sigma T_{\rm eff}(R_{\rm l})^4 \right)^{1/\alpha},
  \end{split}
\end{equation}
where $g_{\rm c}^z$ is the $z$-component of the gravitational acceleration of the BH at the singular point, $g_{\rm c}^z=GM_{\rm BH}z_{\rm c}/(R_{\rm l}^2+z_{\rm c}^2)^{3/2}$.

Here, we put some assumptions on Eq. \ref{eq-mout}.
In order to estimate the surface area of the launching region,
we assume that the radii at the inner and outer edges are
$R=a R_{\rm l}$ $(0<a <1)$ and $R=b R_{\rm l}$ $(b >1)$,
where $a$ and $b$ are constant regardless of the parameters.
We then obtain $S=\pi(b^2-a^2)R_{\rm l}^2$.
Additionally, we put the following assumptions about the singular point on Eq. \ref{eq-mout}:
(1) the gravitational force is larger than the radiation force due to the electron scattering at the singular point, $g_{\rm c}^z \gg \sigma_{\rm e}\sigma T_{\rm eff}(R_{\rm l})^4/c $;
(2) the singular point is close to the disc surface, $z_{\rm c}\ll R_{\rm l}$
(i.e., $g_{\rm c}^z \simeq GM_{\rm BH}z_{\rm c}/R_{\rm l}^3$);
(3) the height of the singular point normalized by $R_{\rm S}$ is constant regardless of the parameters, $z_{\rm c}/R_{\rm S}={\rm const}$. 
These approximations are reasonable because the singular point is located at the vicinity of the sonic point (see Eq. 48 in CAK75). In our simulations, the sonic point is close to the disc surface and the height of its location does not significantly depend on the parameters.

In addition, the relation between the launching radius and the effective temperature at that radius is written as
\begin{equation}
  \label{eq-Rl}
  \frac{R_{\rm l}}{R_{\rm S}}=\left(\frac{3c^5}{16\sigma_{\rm e}\sigma G}\right)^{1/3}
  T_{\rm eff}(R_{\rm l})^{-4/3}\dot m_{\rm sup}^{1/3}M_{\rm BH}^{-1/3},
\end{equation}
based on the standard disc model \citep{1973A&A....24..337S}.
Applying the above assumptions and Eq. \ref{eq-Rl} to Eq. \ref{eq-mout},
the mass-loss rate is rewritten as
\begin{equation}
  \label{eq-mout2}
  \begin{split}
  \dot M_{\rm out}=
  &\frac{\pi(b^2-a^2)}{\sigma_{\rm e}v_{\rm th}}  
  \alpha (1-\alpha)^{(1-\alpha)/\alpha}K^{1/\alpha}
  \left(\frac{\sigma_{\rm e}\sigma}{c}\right)^{1/\alpha}\left(\frac{2G}{c^2}\right)^{2/\alpha}\\
  &\times\left(\frac{Gz_{\rm c}}{R_{\rm S}} \right)^{-(1-\alpha)/\alpha}\left(\frac{3c^5}{16\sigma_{\rm e}\sigma G}\right)^{(3-\alpha)/3\alpha}
  Z_\odot^{-(1-\alpha)/\alpha}\\
  &\times\dot m_{\rm sup}^{(3-\alpha)/3\alpha}T_{\rm eff}(R_{\rm l})^{4/3}
  M_{\rm BH}^{4/3}Z^{(1-\alpha)/\alpha}.
  \end{split}
\end{equation}

\begin{figure}
 \begin{center}
  \includegraphics[width=\columnwidth]{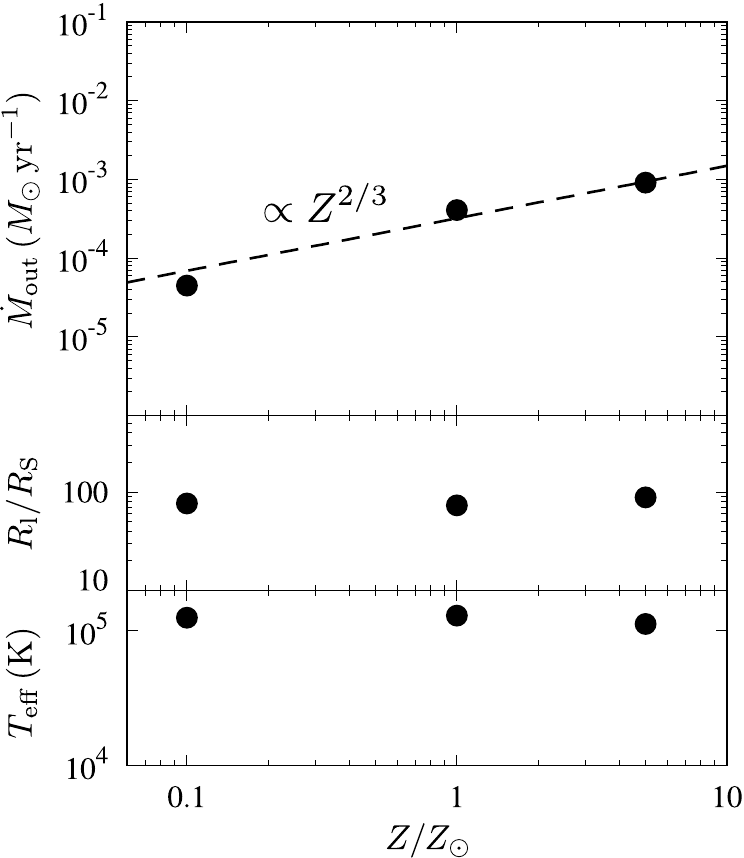}
 \end{center}
 \caption{Mass-loss rate (top panel), launching radius (middle panel)
and effective temperature at the launching radius (bottom panel)
as functions of the metallicity. The BH mass and mass supply rate are fixed ($M_{\rm BH}=10^5\,M_\odot$ and $\dot m_{\rm sup}=0.5$).
The dashed line shows the relation $\dot M_{\rm out}\propto Z^{2/3}$ obtained from 
the semi-analytical model.
 }
 \label{fig4}
\end{figure}

We now compare the semi-analytical model to the results of simulations.
Here, we focus on the $Z$- and $M_{\rm BH}$-dependencies of the mass-loss rate without detailed estimation of $a$, $b$, and $z_{\rm c}$.
Fig. \ref{fig4} shows the $Z$-dependence of the wind properties.
For fixed BH mass ($M_{\rm BH}=10^5\,M_\odot$) and mass supply rate ($\dot m_{\rm sup}=0.5$),
the launching radius and corresponding the effective temperature are almost constant at $R_{\rm l}\sim 100R_{\rm S}$ and $T_{\rm eff}(R_{\rm l})\sim 10^5\,{\rm K}$ (middle and bottom panels). Considering that the effective temperature at the launching radius is constant, we can derive the $Z$-dependence of the mass-loss rate from the semi-analytical model (Eq. \ref{eq-mout2}) as $\dot M_{\rm out}\propto Z^{2/3}$.
In the top panel, the filled circles show the mass-loss rate calculated by our simulations, 
which is well explained by the semi-analytical prediction (dashed line). Thus, we can conclude that the metallicity dependence of the mass-loss rate in the present simulations comes from the $Z$-dependence of the force multiplier in the launching region where the ionization parameter is low (Eq. \ref{eq-fm}).
We note that the absolute value of $\dot M_{\rm out}$
(an intercept of the dashed line in the top panel)
is estimated by fitting the simulation data (filled circles).

\begin{figure}
 \begin{center}
  \includegraphics[width=\columnwidth]{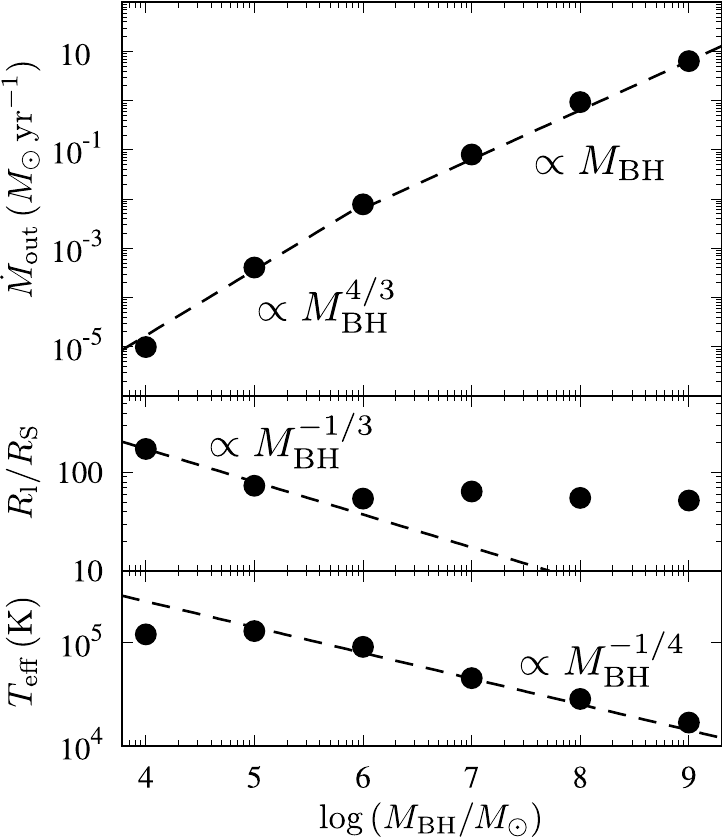}
 \end{center}
 \caption{Mass-loss rate (top panel), launching radius (middle panel)
and effective temperature at the launching radius (bottom panel)
as functions of the BH mass. 
The metallicity and the normalized mass supply rate are fixed ($Z=Z_\odot$ and $\dot m_{\rm sup}=0.5$).
The dashed lines in the top panel
show the lines of $\dot M_{\rm out}\propto M_{\rm BH}^{4/3}$ for $M_{\rm BH} \leq 10^6\,M_\odot$ and $\dot M_{\rm out}\propto M_{\rm BH}$ for $10^6\,M_\odot \leq M_{\rm BH}$.
The dashed lines in the middle and bottom panels indicate the relations of
$R_{\rm l}/R_{\rm S}\propto M_{\rm BH}^{-1/3}$
and $T_{\rm eff}(R_{\rm l}) \propto M_{\rm BH}^{-1/4}$
.}
 \label{fig5}
\end{figure}

The $M_{\rm BH}$-dependence of the wind properties is shown by Fig. \ref{fig5}.
Here, we employ fixed metallicity ($Z=Z_\odot$)
and mass supply rate ($\dot m_{\rm sup}=0.5$).
Note that the choice of computational domain is different for $M\leq 10^6\,M_\odot$ or for $M\leq 10^7\,M_\odot$ as mentioned in Section \ref{sec:result}.
For $M_{\rm BH}\leq 10^6\,M_\odot$, the launching radius of the wind decreases with the growth of the BH mass (middle panel), and the effective temperature at the launching radius is almost constant, $T_{\rm eff}(R_{\rm l})\sim 10^5\,{\rm K}$ (bottom panel). 
We found that these results do not depend on the choice of the computational domain.
This relation can be understood by Eq. \ref{eq-Rl} indicating the $M_{\rm BH}$-dependence of the launching radius, $R_{\rm l}/R_{\rm S}\propto M_{\rm BH}^{-1/3}$, for a constant effective temperature (see the dashed line in the middle panel). In the semi-analytical model (Eq. \ref{eq-mout2}),
the constant effective temperature at the launching radius leads to the $M_{\rm BH}$-dependence of the mass-loss rate of $\dot M_{\rm out}\propto M_{\rm BH}^{4/3}$.
The mass-loss rate calculated by our simulations (filled circles in the top panel) is well fitted by the semi-analytical relation of $\dot M_{\rm out}\propto M_{\rm BH}^{4/3}$ (dashed line in the top panel).
This dependence is determined by the surface area of the launching region, $S\propto R_{\rm l}^2\propto M_{\rm BH}^{4/3}$. 
With the growth of the BH, the extent of the wind launching region, which corresponds to the UV-bright region in the disc, increases, thereby resulting in massive disc winds.

In contrast, for SMBHs ($M_{\rm BH}\geq 10^6\,M_\odot$),
the launching radius is almost constant at
$R_{\rm l}\sim 60R_{\rm S}$ (middle panel),
and the effective temperature at the launching radius decreases
with increasing BH mass (bottom panel).
This is consistent with Eq. \ref{eq-Rl}, 
where we obtain $T_{\rm eff}(R_{\rm l})\propto M_{\rm BH}^{-1/4}$
if $R_{\rm l}/R_{\rm S}$ has no $M_{\rm BH}$-dependence
(see the dashed line in the bottom panel).
Substituting the $M_{\rm BH}$-dependence of the effective temperature
into Eq. \ref{eq-mout2}, the $M_{\rm BH}$-dependence of the mass-loss rate
becomes $\dot M_{\rm out}\propto M_{\rm BH}$ in the semi-analytical model.
This prediction (dashed line in the top panel) well reproduces
the mass-loss rate calculated by our simulations
(filled circles in the top panel). 

The difference of the $M_{\rm BH}$-dependence of the mass loss rate between the intermediate mass range ($\dot M_{\rm out}\propto M_{\rm BH}^{4/3}$)
and the larger mass range ($\dot M_{\rm out}\propto M_{\rm BH}$)
comes from the $M_{\rm BH}$-dependence of the effective temperature at the wind launching radius.
For IMBHs, the disc wind is ejected from the UV-bright region so that the effective temperature at the launching radius is around $T_{\rm eff}\sim 10^5\,{\rm K}$.
The disc temperature becomes law and the UV-bright region shifts inwards as the BH mass increases. 
For SMBHs, the most of the UV bright region is 
located within $r_{\rm in}=30R_{\rm S}$. 
In this case, 
the wind blows from near $r_{\rm in}$ or slightly outside $r_{\rm in}$,
where the effective temperature is smaller than $T_{\rm eff}\sim 10^5\,{\rm K}$.

In our simulations, we set the inner radius of the computational domain to be $30R_{\rm S}$, which corresponds to the assumption that the corona is present inside of $\sim\!30R_{\rm S}$. 
As we have mentioned above,
the size of coronal region is still under debate but is expected to be in the range $10-100R_{\rm S}$ \citep[e.g.,][]{2016A&A...594A..71E,2018MNRAS.480.1247K}, which is not far from our setup. Thus, the result that the $M_{\rm BH}$-dependence changes at around $M_{\rm BH}=10^6\,M_\odot$ would not largely affected even if we set the corona size to be different value.
The simulations of the disc winds with more realistic corona model are left for future works.

\section{Summary and discussions}
\label{sec:summary}
By way of radiation hydrodynamics simulations, we have studied the metallicity and BH mass dependencies of the line-driven winds from AGN discs. 

Our findings can be summarized as follows:
\begin{enumerate}
\item Denser and faster disc winds are launched toward upper, i.e., nearly vertical,  directions for higher metallicity and larger BH mass.
\item The line-driven winds effectively reduce the mass accretion rate for the BH mass $M_{\rm BH}\gtrsim 10^5\,M_\odot$ in high-metallicity environments.
When the metallicity is $0.1Z_\odot$, the mass accretion rate is comparable to the mass supply rate regardless of the BH mass, but when the metallicity is $Z_\odot$ ($5Z_\odot$), the mass accretion rate is less than $\sim\!50$ per cent of the mass supply rate for $M_{\rm BH}\gtrsim 10^6\,M_\odot$ ($M_{\rm BH}\gtrsim 10^5\,M_\odot$, respectively).
  
\item Suppression of the mass accretion due to the line-driven winds is more significant for the SMBH range ($\ga 10^6M_{\sun}$) than for the IMBH range ($\la 10^6M_{\sun}$) unless the mass supply rate is very low ($\dot m_{\rm sup} \la 0.1$). 
\item A semi-analytical model based on the stellar wind model of CAK75 reproduces well the metallicity dependence and BH mass dependence of our wind mass-loss rate. 
The metallicity dependence comes from the metallicity dependence of the force multiplier in the (low-ionized) wind launching region, while the BH mass dependence is explained by the BH mass dependence of the surface area of the launching region and of the effective temperature at the launching radius.
\end{enumerate}

\begin{figure}
 \begin{center}
  \includegraphics[width=\columnwidth]{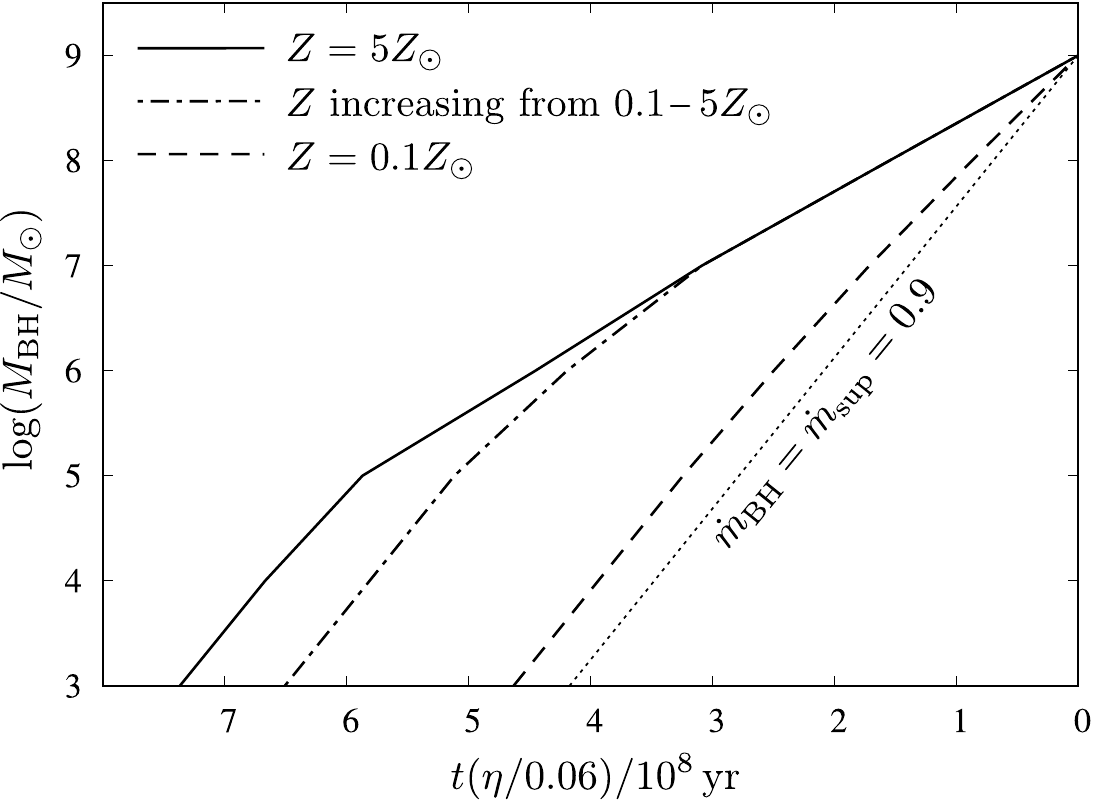}
 \end{center}
 \caption{
 Accretion growth of BH mass as a function of a look-back time calculated from $M_{\rm BH}=10^9\,M_\odot$.
 The dotted line shows the unhindered growth at $\dot m_{\rm BH}=0.9$. The dashed, dotted-dashed, and solid lines show the growths with the disc winds for $Z=0.1Z_\odot$, $Z$ increasing with the BH mass ($Z=0.1Z_\odot$ for $10^3\,M_\odot\leq M_{\rm BH}<10^{5}\,M_\odot$,
$Z=Z_\odot$ for $10^5\,M_\odot\leq M_{\rm BH}<10^7\,M_\odot$,
and $Z=5Z_\odot$ for $10^7\,M_\odot\leq M_{\rm BH}<10^9\,M_\odot$), and $Z=5Z_\odot$. The energy conversion rate $\eta$ is set to 0.06 consistent with the numerical simulations.
}
 \label{fig6}
\end{figure}

The growth of BHs is effectively suppressed 
in the case where the metallicity is high ($Z=5Z_\odot$) 
at all the time
or increases from $0.1Z_\odot$ to $5Z_\odot$ with the growth of BH mass.
We show the growth time of BHs considering the wind mass loss estimated by our simulations.
Here we focus on the case of $\dot m_{\rm sup}=0.9$, which corresponds to relatively rapid growth within the sub-Eddington regime.
The time needed for a BH with mass $M_{\rm BH}$ to grow to $10^9\,M_\odot$ is estimated as
\begin{equation}
  t=\frac{\eta \sigma_{\rm e} c}{4\pi G } \int_{M_{\rm BH}}^{10^9\,M_\odot}\frac{dM_{\rm BH}^\prime}{\dot m_{\rm BH}(M_{\rm BH}^\prime, Z)M_{\rm BH}^\prime},  
\end{equation}  
where the normalized mass accretion rate $\dot m_{\rm BH}(M_{\rm BH},\, Z)$ 
is given by the results of our simulations.
This ratio is assumed to be constant in each dex in the BH mass: $\dot m_{\rm BH}(M_{\rm BH},\, Z)=\dot m_{\rm BH}(10^n\,M_\odot,\, Z)$ for $10^n\,M_\odot\leq M_{\rm BH}<10^{n+1}\,M_\odot$ with $n=3,\,4,\,\ldots,\,8$.
Fig. \ref{fig6} shows the accretion growth of the BH mass calculated in this way for three cases with different metallicities.
The dotted line shows the unhindered growth at the given supply rate with the constant Eddington ratio $\dot m_{\rm BH}=\dot m_{\rm sup}=0.9$.
If the metallicity is kept constant at $Z=0.1Z_\odot$ all the time (dashed line), 
the growth time is not so different from that for the unhindered growth (dotted line) because of 
a small amount of mass loss due to the disc wind. 
Another more plausible case 
considered here is that where the metallicity gradually increases from $0.1Z_\odot$ to $5Z_\odot$ ($Z=0.1Z_\odot$ for $10^3\,M_\odot\leq M_{\rm BH}<10^{5}\,M_\odot$,
$Z=Z_\odot$ for $10^5\,M_\odot\leq M_{\rm BH}<10^7\,M_\odot$,
and $Z=5Z_\odot$ for $10^7\,M_\odot\leq M_{\rm BH}<10^9\,M_\odot$). 
In this case, the growth rate is close to that for $\dot m_{\rm BH}=0.9$ until $M_{\rm BH}\sim 10^5\, M_\odot$,
but it gradually slows down for $M_{\rm BH}\geq 10^6\,M_\odot$ (dashed-dotted line).
The look-back times from $M_{\rm BH}=10^9\,M_\odot$ to
$10^5\,M_\odot$, $10^4\,M_\odot$, and $10^3\,M_\odot$ are
$\sim\! 5.1\times 10^8\,{\rm yr}$, $\sim\! 5.8\times 10^8\,{\rm yr}$, and $\sim\! 6.5\times 10^8\,{\rm yr}$,
which are $\sim\!1.6$--$1.8$ times larger than those for $\dot m_{\rm BH}=0.9$.
In the high-metallicity case in which the metallicity is $Z=5Z_\odot$ in the entire range of the BH mass (solid line),
the growth rate for $M_{\rm BH}\lesssim 10^5\,M_\odot$ 
is similar to those for the other three cases
because even in the high-metallicity environment,
the small surface area of wind launching, i.e., UV-bright, region results in weaker wind mass loss (see Section \ref{sec:reason}).
The BH growth slows down at $M_{\rm BH}=10^5\,M_\odot$
because massive disc winds emerge for higher BH mass. 
The growth times from $10^5\,M_\odot$, $10^4\,M_\odot$, and $10^3\,M_\odot$ BHs are 
$\sim\! 5.9\times 10^8\,{\rm yr}$, $\sim\! 6.7\times 10^8\,{\rm yr}$, and $\sim\! 7.4\times 10^8\,{\rm yr}$,
respectively, which are $\sim\!1.8$--$2.1$ times larger than those of the unhindered growth. 
The results above indicate that the line-driven wind does not significantly affect early growth of the BHs in the range $M_{\rm BH}\lesssim 10^5\,M_\odot$,
while it makes
the growth timescale in the later (i.e., more massive) phases about twice longer unless the metallicity remains sub-solar $\sim\! 0.1Z_{\sun}$.
Note that with growth time twice longer than in the unhindered case, in order to reach $10^{9}M_{\sun}$ during the time interval in which a $10^3M_{\sun}$ seed BH grows to a $10^9M_{\sun}$ SMBH in the unhindered growth, the seed BH must be as massive as $10^{6}M_{\sun}$. 

\begin{figure}
 \begin{center}
  \includegraphics[width=\columnwidth]{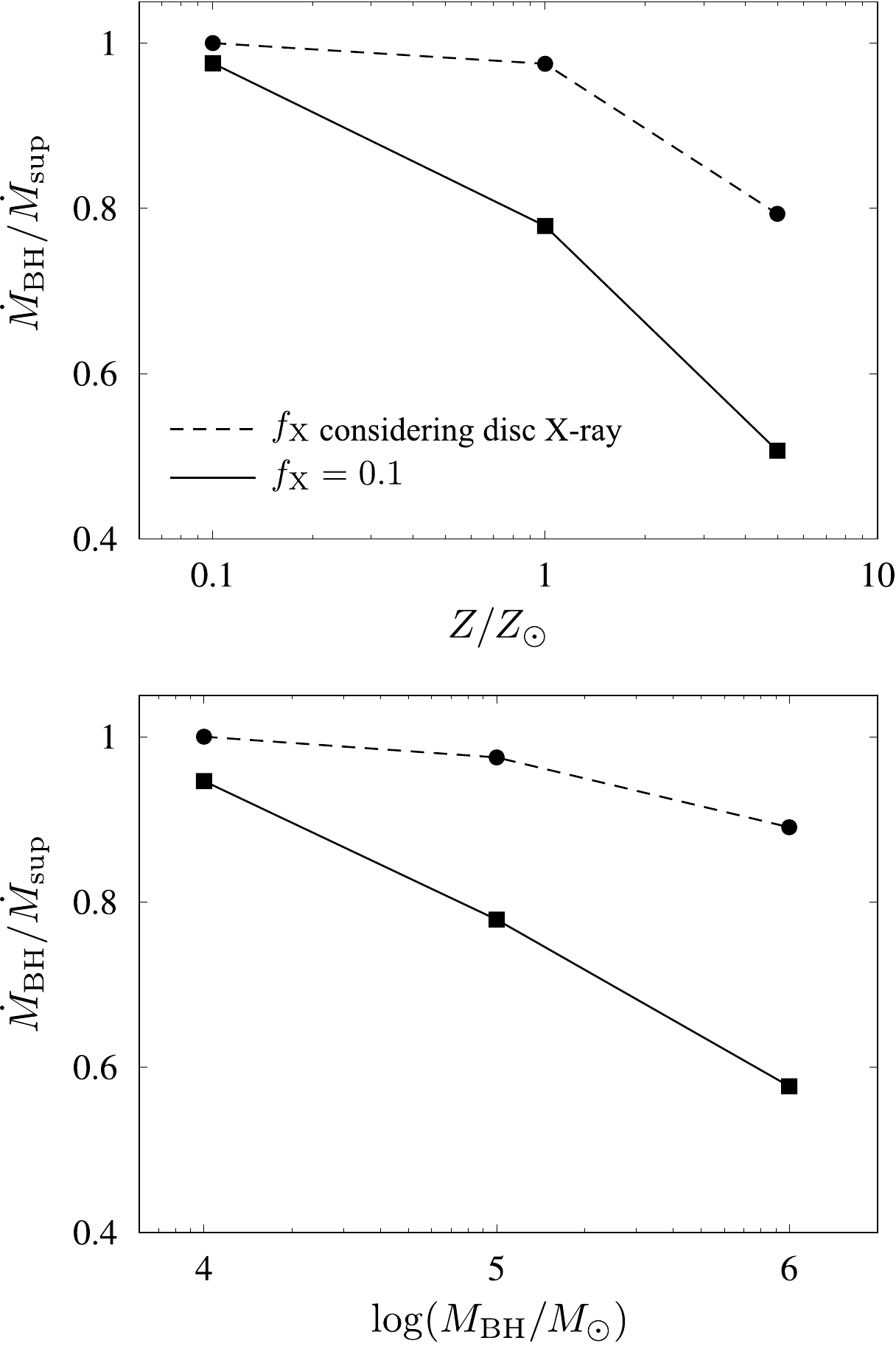}
 \end{center}
 \caption{Mass accretion rate normalized by the mass supply rate as a function of the metallicty (top panel) and the BH mass (bottom panel).
   In both panels, the mass supply rate is $\dot m_{\rm sup}=0.5$.
   In the top panel, the BH mass is set to $M_{\rm BH}=10^5M_\odot$.
   In the bottom panel, the metallicity is set to $Z=Z_\odot$.
   Dashed lines show the normalized mass accretion rates employing $f_{\rm X}=0.94$, 0.89, and 0.78 for $M_{\rm BH}=10^4\,M_\odot$, $10^5\,M_\odot$, and
   $10^6\,M_\odot$.
   Solid lines show the results using $f_{\rm X}=0.1$.
}
 \label{fig7}
\end{figure}

We have investigated the line-driven winds in a wide range of BH masses by using simple setup ignoring the X-rays emitted from the disc. However, for the IMBHs, the accretion disc is also bright in the X-rays.
Strong X-ray irradiation overionizes the metals 
and reduces the opacity of bound-bound transitions in the UV band,
and thus the line force becomes powerless \citep{1990ApJ...365..321S}.
When $\dot m_{\rm sup}=0.5$,
while the line-driving UV luminosities ($200$--$3200$\,\AA) are 8 per cent, 17 per cent, 
and 33 per cent of the disc luminosities,
the ionizing X-ray luminosities integrated across 
$40\,{\rm eV}$--$13.6\,{\rm keV}$
become as high as 94 per cent, 89 per cent, and 78 per cent of the disc luminosities for
$M_{\rm BH}=10^4\,M_\odot$, $10^5\,M_\odot$, and $10^6\,M_\odot$, respectively.
Here we study the effects of X-rays on the mass accretion rate by modifying the value of $f_{\rm X}$, 
assuming that X-rays are emitted only from the central point source.  
Although this assumption might be somewhat too simple,
the following results would help us understand X-ray effect on mass accretion onto IMBHs.
Fig. \ref{fig7} shows that $\dot M_{\rm BH}/\dot M_{\rm sup}$ for modified $f_{\rm X}$ (dashed line) 
is larger than that for our standard setup of $f_{\rm X}=0.1$ (solid line).
When the mass supply rate and the BH mass are set to
$\dot m_{\rm sup}=0.5$ and $M_{\rm BH}=10^5\,M_\odot$ (top panel),
the normalized mass accretion rate calculated with $f_{\rm X}=0.89$
is close to 1 for $Z\leq Z_\odot$ and
still larger ($\dot M_{\rm BH}/\dot M_{\rm sup}\sim 0.8$) 
than that for $f_{\rm X}=0.1$ ($\dot M_{\rm BH}/\dot M_{\rm sup}\sim 0.5$) even for $Z=5Z_\odot$.
When we focus on $\dot m_{\rm sup}=0.5$ and $Z=Z_\odot$ (bottom panel),
modified $f_{\rm X}$ is set to $f_{\rm X}=0.94$, 0.89, and 0.78
for $M_{\rm BH}=10^4\,M_\odot$, $10^5\,M_\odot$, and $10^6\,M_\odot$, respectively.
The mass accretion rate is comparable to the mass supply rate for $M_{\rm BH}\leq 10^5\,M_\odot$
and $\sim\!80$ per cent of the mass supply rate even for $M_{\rm BH}=10^6\,M_\odot$.
This is because the ionization due to the strong X-ray suppresses the line force and reduces the mass-loss rate of the line-driven winds.
These results indicate that the disc X-ray has a tendency to reduce the mass-loss rate of the line-driven winds for the IMBHs.
We note that the effect of the X-ray cannot be quantified accurately unless we consider the spectral energy distributions and the geometries of the source. The results are also affected by whether the IMBH has a X-ray corona like an AGN. 
Additionally, 
it is important to take into account disc winds accelerated by other mechanisms such as magnetic forces in order to investigate more realistic accretion processes for the IMBHs.

In addition, there are several limitations in the current treatment of radiation transfer. 
We have employed the mass extinction coefficient for the X-rays changing abruptly from $\sigma_{\rm X}=\sigma_{\rm e}$ ($\xi >10^5$) to $100\sigma_{\rm e }$ ($\xi <10^5$), corresponding the assumption that the X-rays are heavily attenuated below $\xi \simeq 10^5$.
In reality, also in the range $10^3\lesssim \xi \lesssim 10^5$, there should be some X-ray attenuation and our treatment might underestimate this effect. 
For the attenuation of the line-driving UV radiation, we have used the opacity $\sigma_{\rm UV}=\sigma_{\rm e}$. 
More correct treatment should include
the effect of line overlapping, i.e., a part of UV radiation is absorbed by the wind material by intervening lines.
In N20, we assessed the impact of different $\sigma_{\rm X}$ by comparing the case with the same step-function-like $\sigma_{\rm X}$ as in our fiducial model here and the case with $\sigma_{\rm X}=\sigma_{\rm e}$ for all $\xi$ but with the UV opacity $\sigma_{\rm UV}=0$ \citep[see also][]{2004ApJ...616..688P}.
We found that the mass loss rate is a factor of $\sim\!2-4$ smaller in the latter case than in the former case (see Fig.5 in N20 for details). 
To construct a realistic wind model, 
more sophisticated treatment of the opacity for X-ray and line-driving radiation would be required.
Also, we do not treat possible metallicity dependence of the X-ray opacity, which is expected to be proportional to the metallicity \citep[][]{2011piim.book.....D}.
To investigate the consequence of this treatment, we have calculated two additional runs by changing the X-ray opacity at $\xi<10^5$ in proportion to the metallicity and found that the accretion rate hardly changes in both cases. 
At metallicity $Z=0.1Z_{\odot}$, we 
have studied the case of ($M_{\rm BH}=10^5\,M_\odot$, $\dot m_{\rm sup}=0.5$) by reducing $\sigma_{\rm X}$ by a factor of ten ($\sigma_{\rm X}=10\sigma_{\rm e}$) at $\xi<10^5$. 
Now, due to less attenuated X-rays, the gas is more ionized and no wind is launched, i.e., $\dot M_{\rm BH}=\dot M_{\rm sup}$. 
At such low metallicity, the line force is small and we found 
$\dot M_{\rm BH} = 0.98\dot M_{\rm sup}$
in our fiducial case anyway for the same set of the parameters. 
When $Z=5Z_\odot$, we examined the case of ($M_{\rm BH}=10^5\,M_\odot$, $\dot m_{\rm sup}=0.5$) and found that the mass accretion rate does not change even with the X-ray opacity enhanced by a factor of five ($\dot M_{\rm BH} = 0.51\dot M_{\rm sup}$ for $\sigma_{\rm X}=100\sigma_{\rm e}$ and $\dot M_{\rm BH} = 0.50\dot M_{\rm sup}$ for $\sigma_{\rm X}=500\sigma_{\rm e}$).
This is because the X-ray are already totally attenuated even in the fiducial case of $\sigma_{\rm X}=100\sigma_{\rm e}$.
More detailed modelling of the X-ray opacity is desirable in future works in this field.

The scattered and reprocessed photons are ignored in our model.
Based on post-processed radiation transfer calculations,
\citet{2010MNRAS.408.1396S} and 
\citet{ 2014ApJ...789...19H} 
reported that these secondary photons ionize the material and weaken the line force.
In order to assess these effects due to the secondary photons, hydrodynamics simulations coupled with the radiation transfer including scattered and reprocessed radiations are needed.
Such simulations impose too much computational cost at present, but will be important future works.

Our model does not solve the structure of the accretion disc itself. In the current method, the geometrically thin and optically thick disc lies below the computational domain. The disc surface is located at the $\theta=90^\circ$ boundary. The disc is treated as an external radiation source and the photons are supposed to be steadily emitted from the vicinity of the equatorial plane of the disc. 
Although the reduction of the mass accretion rate of the disc via the launching the wind is taken into consideration, self-consistent simulations of the wind and disc structure would be necessary to understand the detailed accretion processes around the BH. 

In this paper, we focus on the sub-Eddington regime, but the accretion processes in the super-Eddington sources are important to understand the rapidly growing BHs.
Although the line force was not considered, radiation hydrodynamics simulations of super-Eddington accretion flow found that the radiation pressure on the electrons accelerates the disc winds \citep[e.g.,][]{2009PASJ...61L...7O, 2011ApJ...736....2O, 2018PASJ...70...22K}. 
Even in the super-Eddington sources, the line force would be effective in a low-temperature region outside the photon trapping radius. In addition to the continuum-driven wind (i.e., accelerated by the radiation pressure on electrons) near the BH, the line-driven wind launched from the outer region of the disc may affect the accretion growth of the BH in the super-Eddington sources. The simulations considering such situations are left as future works.

\section*{Acknowledgements}
The authors would like to thank to Kazuyuki Sugimura for useful discussions. Numerical computations were carried out on Cray XC50 at Center for Computational Astrophysics, National Astronomical Observatory of Japan and on the computer cluster, 
\verb'Draco', at Frontier Research Institute for Interdisciplinary Sciences of Tohoku University. 
This work was supported by JSPS KAKENHI Grant Numbers JP20K14525, JP20H00178 (MN), JP17H01102, JP17H06360, JP17H02869 (K. Omukai), JP18K03710, JP21H04488 (K. Ohsuga), by MEXT as "Program for Promoting Researches on the Supercomputer Fugaku" (Toward a unified view of the universe: from large scale structures to planets, K. Ohsuga), and by Joint Institute for Computational Fundamental Science (JICFuS, K. Ohsuga).

\section*{Data Availability}
The data underlying this article will be shared on reasonable request to the corresponding author.



\bibliographystyle{mnras}
\bibliography{ref} 








\bsp	
\label{lastpage}
\end{document}